
\input amstex
\documentstyle{amsppt}
\null
\magnification=\magstep1
\hsize=15.2truecm
\vsize=22truecm
\voffset=0\baselineskip
\par
\noindent
\tolerance=10000
\hfill{gr-qc/9501012}
\vskip.2in
\centerline{\bf NONSTATIONARY KERR CONGRUENCES}
\vskip.3in
\centerline{\bf A. Burinskii$^*$, R. P. Kerr$^{**}$}
\vskip .2in
 \+&\centerline{\it $^*$Nuclear Safety Institute, Russian
Academy of Sciences}\cr
\+&\centerline{\it B. Tulskaya 52, Moscow 113191, Russia; e-mail:
grg\@ibrae.msk.su}\cr
\+&\centerline{\it $^{**}$Department of Mathematics, University
 of Canterbury, }\cr
\+&\centerline{\it Private Bag, Christchurch 1, New Zealand}\cr
\vskip.5in \midinsert \baselineskip=10pt plus.1pt {\bf
ABSTRACT}\par \vskip.1in
The Kerr solution is defined by a null congruence which is geodesic
and shear free and has a singular line  contained in a bounded region of
space.
A generalization of the Kerr congruence for a  nonstationary case is obtained.
We find a  nonstationary shear free geodesic null congruence which is
generated by a given analytical complex world line.  Solutions of the
Einstein equations are analyzed. It is shown that there exists complex
radiative solution which is generalization of the Kerr solution and the
Kinnersley accelerating solution for "photon rocket".
\endinsert
\baselineskip=17pt
\bigskip
\def\b{\bar}
\def\d{\partial}
\def\D{\Delta}

\def\G{\Gamma}
\def\l{\lambda}

\def\m{\mu}
\def\n{\nu}
\def\p{\psi}
\def\q{\b q}

\def\t{\tau}
\def\x{\phi}

\def\~{\tilde}
\def\h{\eta}
\def\z{\zeta}
\def\Z{{\b\zeta}}
\def\Y{{\b Y}}
\def\cZ{{\b Z}}
\def\`{\dot}
{\bf 1. INTRODUCTION}
\par
\bigskip
      The algebraically special solutions of the Einstein and
Einstein-Maxwell equations are characterized by the existence of geodesic
shear free principal null congruences. Setting up the  congruence is a first
step to solving the corresponding field equations.  The Kerr theorem [1,2,3]
gives a rule how to construct such congruences:
An arbitrary  geodesic shear-free null congruence in
Minkowski space is defined by a function $Y$ which is a solution of the
equation
$$         F  = 0 ,                            \eqno(1.1) $$
where   $F (\l_1,\l_2,Y)$   is an arbitrary analytic
function of the projective twistor coordinates
$$ \l_1 = \z - Y v, \qquad \l_2 =u + Y \Z, \qquad Y\ .\eqno(1.2)$$
  In this paper  we consider Kerr-Schild space-times containing geodesic
and shear free congruences.
First we analyze one of the two principal null congruences
of the Kerr   metric.
It belongs to a class of congruences for which the singularities
are contained in a bounded region of space [4,5] .  In this case the
equation $ F=0$ may be solved in explicit form. We find a representation
for the function F in which the congruence is defined by a source moving in
complex Minkowski space $CM^4$ along a straight complex world line.
This representation reproduces, in the Kerr-Schild formalism,
a retarded time construction which was considered
before by Lind and Newman [6,7].
\par
      Second, we obtain a nonstationary generalization of the principal null
congruence of the Kerr metric, in which the congruence  is generated
by a given complex world line $x_0(\t)$ depending on a complex time $\t$. We
find also the necessary conditions under which the congruence satisfies the
Kerr theorem:
\par
  i) the world-line $ x_0(\t)$ must have an analytical dependence on
 the complex time $ \t,$
\par
  ii) $\t$ must be a "left" solution of the retarded
 time equation, corresponding to the "left" null plane of the
 complex light cone.
\par
In the last section we analyze the field equations and find the complex
solutions in which the
metric is not vacuum but has a
 Ricci tensor corresponding the stress tensor of null radiation. This
solutions contain as a particular case the Kerr solution and the Kinnersley
accelerating solution for a "photon rocket" [3,8].
\par
In the notation we follow the work of Debney, Kerr and Schild [1].
In the four-dimensional space-time with signature  $(+++-)$, the
null tetrad $ e_1, e_2, e_3, e_4 $ is given by
$$g_{ab}=
e_a^\m e_{b\m} =\left(\matrix 0&1&0&0\cr1&0&0&0\cr 0&0&0&1\cr0&0&1&0
\endmatrix \right) = g^{ab}. \eqno(1.3) $$ $e^3, e^4$ are real null
vectors, $ e^1, e^2 $ are complex conjugates. The Ricci rotation
coefficients are $$\G ^a_{bc} = - \quad e^a_{\m;\n} e_b^\m e_c^\n.
\eqno(1.4) $$ Let the principal null congruence which is being considered
have the  $e^3$  direction as tangent.
It will be geodesic if and only if $\G_{424} = 0$ and shear
free if and only if $\G_{422} = 0$.
The corresponding complex conjugate terms are
$\G_{414} = 0$ and $\G_{411} = 0$.  The Kerr-Schild metric $$g_{\m\n} =
\h_{\m\n} + 2 h e^3_{\m} e^3_{\n} \eqno(1.5) $$ has the property that the
principal null direction $ e^3 $ is also a null direction with respect to
the auxiliary Minkowski space with metric
$$\h = dx^2 + dy^2 + dz^2 -dt^2 = 2du dv+2d\Z d\z \ \eqno(1.6) $$
where the null coordinates are related to Cartesian coordinates by
$$\eqalign{2^{1\over2}\z &= x+iy ,\qquad 2^{1\over2} \Z = x-iy ,\cr
2^{1\over2}u & = z + t ,\qquad 2^{1\over2}v = z - t}  \eqno (1.7) $$
The field of the null directions $e^3$ is defined by
the complex function $Y $ $$ e^3 = du+ \Y d \z  + Y d \Z - Y \Y d v.
\eqno(1.8) $$ Complete the null tetrad $e_a^\m$ is as follows: $$\eqalign{
e^1 &= d \z - Y d v; \cr e^2 &= d \Z - \Y d v; \cr e^4 &=  d v - h e^3. }
   \eqno(1.9) $$ The inverse tetrad has the form $$\eqalign{ \d_1 &= \d_\z
 - \Y \d_u ; \cr \d_2 &=  \d_\Z - Y \d_u ; \cr \d_3 &=  \d_u - h \d_4 ;
\cr \d_4 &=  \d_v + Y \d_\z + \Y \d_\Z - Y  \Y \d_u .  } \eqno(1.10) $$ It
was shown in [1] that $$ \G_{42} = \G_{42a} e^a  = - d Y - h Y,_4 e^4 .
\eqno (1.11) $$ The congruence  $e^3 $ is geodesic if $ \G_{424} =
-Y,_4 (1-h) = 0, $
and is shear free if $ \G_{422} = -Y,_2 = 0.$
Thus  the function $ Y $ with conditions
 $$ Y,_2 = Y,_4 = 0,  \eqno (1.12) $$
defines a shear free geodesic congruence.
\baselineskip=17pt
\bigskip
{\bf 2.THE KERR THEOREM FOR THE GENERAL CONGRUENCES}\par
\bigskip
Given the geodesic and shear free nature (1.12) of the null congruence,
the Kerr theorem [1,2,3], (1.1) and (1.2) holds.
      Following [1], we consider the differential of
the function $Y$  in the case of $Y,_2 =  Y,_4 = 0 :$
      $$ d Y = Y,_a e^a  = Y,_1 e^1 + Y,_3 e^3.
     \eqno(2.1)$$
 $ Z = Y,_1$ is the complex expansion of the null congruence
(expansion + $ i$ rotation).
As the first step we work out the form of $Y,_3$. By using relations
(1.10) and their commutators we find
  $$   Z,_2 = (Z - \cZ) Y,_3.              \eqno (2.2) $$
Straightforward differentiation of $Y,_3$ gives the equation
$$ Y,_{32} = ( Y,_3) ^2    .           \eqno (2.3) $$
And by using (2.2) and (2.3) we obtain the equation
  $$ (Z^{-1} Y,_3),_2 = \cZ ( Z^{-1} Y,_3 )^2.  \eqno (2.4)$$
This is a first-order differential equation for the function
$Z^{-1} Y,_3$.
The general solution
has the form
$$  Y,_3 = Z ( \x - \Y) ^{-1},\eqno(2.5)$$
where $\x$ is an arbitrary solution of the equation $$ \x,_2 =0.  \eqno
(2.6)$$ Substitution $Y,_3$ in (2.1) implies $$Z^{-1} (\Y - \x) d Y = \x
(d \z -Y d v) + (du + Y d \Z). \eqno (2.7)$$
Differentiating the function $F (\l_1,\l_2,Y)$
 and comparing the result with $(2.7)$ we find that $$ PZ^{-1}= - \quad
 d F / d Y  , \qquad P = \d_{\l_1} F - \Y \d_{\l_2} F, \qquad
\m = \d_{\l_2} F, \eqno(2.8) $$
where the function $P$ is defined
$$P = \m (\x - \b Y).  \eqno(2.9) $$
The singular region of the congruence where the complex divergence
$Z$ blows up is defined by the system of equations $$ F=0, \qquad
d F / d Y =0.  \eqno(2.10) $$
\baselineskip=17pt
\bigskip {\bf 3.  STATIONARY CONGRUENCES HAVING
SINGULARITIES CONTAINED IN  A BOUNDED REGION}\par \bigskip
The null congruence with tangent $ e^3 $ is said to be stationary if
there exists a real timelike vectorfield K such that $ K Y = 0 $.
The stationary congruences having singularities contained in
a bounded region have been considered in papers $[4,5]$.
In this case function $F$ must be at most quadratic in $Y$,
$$F \equiv a_0 +a_1 Y + a_2 Y^2 + (q Y + c) \l_1 - (p Y + \q) \l_2,
\eqno(3.1) $$
where coefficients $ c$ and $ p$ are real constants
and $a_0, a_1, a_2,  q, \q, $  are complex constants.
The solutions of the equation $ F = 0$ and the equations for the
singularities may be found in explicit form. An analysis shows that this
case gives the Kerr solution up to the Poincare group of motions.
\par
The form (3.1) of $ F$ implies that $$K Y = 0.  \eqno(3.2)$$
where $$ K= c \d_u + \bar q \d_\z + q \d_\Z - p \d_v . \eqno(3.2a)$$
\par
There is another equivalent form of  $ F$ which allows one to introduce
a retarded time parameter.  One can represent (3.1) in the form
$$ F \equiv (\l_1 - \l_1^0) K \l_2 - (\l_2 -\l_2^0) K \l_1.
\eqno(3.3) $$
Here we introduce a complex world line  $ x_0^\m (\t)$
parametrized by the complex time parameter $\t$. The
coordinates of this world line are complex,
$ x_0 (\t)= (\z_0, \Z_0, u_0 ,v_0) \in CM^4.$
Thus $\Z_0$ and $\z_0$ are not necessarily complex conjugates.
The vector $K$ may be expressed now in the form
$$ K(\t) = \`x_0^\m(\t) \d_\m, \eqno(3.4)$$
where the dot denotes $\d_\tau$.
\par
  In this section we consider a straight world line with
3-velocity $\b v$ in $CM^4$, for which the Kerr-Schild map
(1.5) gives the Kerr solution:
$$ x_0^\m (\t) = x_0^\m (0) + \xi^\m \t; \qquad \xi^\m = (1,\b v), \eqno(3.5)
$$
$K=\xi^\m \d_\m$ is independent of $\t$ and $\xi^\m$ is a Killing vector of
the Kerr  solution.
The parameters in (3.3) are then defined
 $$ \l_1 = \z - Y v, \qquad \l_2 =u + Y  \Z,\eqno(3.6) $$
$$ \l_1^0 (\t)= \z_0(\t) - Y v_0(\t), \qquad
 \l_2^0(\t) =u_0(\t) + Y \Z_0(\t),\eqno(3.7) $$
where the twistor components with zero indices denote the values
on the points of complex world-line
$ x_0 (\t).$
\par
 To obtain the geometrical picture of this representation
we consider a complex light cone with the vertex at some point $x_0$ of the
complex world line $x_0^\m(\t)$. The solutions of the complex light cone
equation $$(x_\m - x_{0 \m})(x^\m -x_0^\m) = 0 ,\eqno(3.8)$$  split into
two families of  null planes:  "left"
planes $$ x_L = x_0(\t) + \alpha e^1 + \beta e^3    , \eqno(L)$$
and "right" planes
$$ x_R = x_0(\t) + \alpha e^2 + \beta e^3, \eqno(R)$$
where $ \alpha$ and $\beta$ are arbitrary parameters
on these planes. Obviously (L) and (R) are solutions of (3.8) with the
metric (1.2).  The twistor parameters $\l_1$ and $\l_2 $ may
be represented in the form $$ \l_1 = x^\m e^1_\m , \qquad \l_2 = x^\m
(e^3_\m - \Y e^1_\m). \eqno(3.9)$$ Substitution of $(L)$ in (3.9)
shows that for an arbitrary Y the twistor parameters $ \l_1$ and $\l_2$
are constants on the "left" planes $$ \l_1 = \l_1^0(\t);\qquad \l_2
= \l_2^0(\t), \eqno (3.10)$$ and the equation $F = 0$ is fulfilled.
One obtains the relations
$$ K\l_1 =\`\z_0 - Y \`v_0 =
\d_\t \l_1^0, \qquad K\l_2 =\`u_0 + Y \`\Z_0 = \d_\t \l_2^0.
\eqno(3.11) $$
In spite of an explicit dependence of the parameters of the $ F$ in (3.3)
on $\t$ via (3.7) this dependence is absent really in consequence of the
relations
$$\l_1^0 (x_0(\t)) = \l_1^0 (x_0(0)) + \t K \l_1,\quad \l_2^0 (x_0(\t)) =
\l_2^0 (x_0(0)) + \t K \l_2,  \eqno (3.12)$$ and cancellation of
the terms proportional to $ \t$.
\par
 Thus (3.1) and (3.3) are equivalent forms of  F. A real cut
of a "left" null plane gives a real null line which is defined by
the twistor coordinates $(\l_1^0(\t), \l_2^0(\t), Y)$.  Thus the
Kerr principal null
congruence arises as the real cut of the family of the "left" null planes
of the complex light cones the vertices of which lie on the straight complex
world line $x_0(\t)$.  By writing the function F in the form $$ F = A
  Y^2 + B Y + C, \eqno (3.13)$$ where  $$\eqalign { A &= (\Z
  - \Z_0) \`v_0 - (v-v_0) \`\Z_0 ;\cr B &= (u-u_0) \`v_0 + (\z - \z_0 ) \`\Z_0
  - (\Z - \Z_0) \`\z_0 - (v - v_0) \`u_0 ;\cr C &= (\z - \z_0 ) \`u_0 - (u -
u_0) \`\z_0, } \eqno(3.14) $$ one can find two explicit solutions for the
function $Y(x)$
 $$ Y_{1,2} = (- B \pm \D )/2A, \eqno(3.15)$$
 where $ \D = (B^2 - 4AC)^{1/2}.$ On the other hand (2.8),(2.9)
and (3.10)
imply $$ Y = - (B + PZ^{-1})/2A, \eqno (3.16) $$ and consequently
$$PZ^{-1} = \mp \D. \eqno (3.17)  $$
This relation reflects a twofoldedness of the Kerr geometry.
The complex radial coordinate $PZ^{-1}$ is related to
the Kerr coordinates [1] by
$PZ^{-1} = r + ia cos \theta $. A change of the sign corresponds to a
transition from the positive $r$ sheet of the metric to the
negative one where $ r \leq 0 $.
By using (2.8), (2.9) and (3.4) we find $$ P = \Y K \l_1 + K \l_2 = \`x_o^\m
(\t) e^3_\m . \eqno (3.18).$$
Like the function $F$ and vector $K$, the coefficients $A,B,C$, as well
as  $Y, P, Z^{-1}$ for the Kerr solution do not depend
on $\t$, in spite of the presence of $ \t$ in the formulas containing
$x_0(\t)$.  Nevertheless this parameter $\t$ may be defined for each point
$ x$
of the Kerr space-time and plays really the role of a retarded time parameter.
Its value for a given point $x$ may be defined by using the solution $Y(x)
$ and by forming the twistor parameters (2.8) which fix the
"left" null plane (3.10). A point of intersection of this plane
with the complex world-line $x_0(\t)$ gives a value of the "left"
retarded time $\t_L$.  Thus $\t_L$ is in fact a complex scalar function on
the space-time $\t_L(x)$.
Since $K x^\m = \dot x_0^\m$, action $K$ on the light cone equation
$(x_\m - x_{0 \m}(\t_L))(x^\m -x_0^\m(\t_L)) = 0 $ yields
$$ K(\t) \t_L(x) = 1 , \eqno (3.19)$$ and then by using (3.10) one finds that
\footnote{ The projective twistor parameters $ \l_1, \l_2, Y $ can be written
in twistor notation  $ Z = (\m ^A , \p _{\dot A}),$
$\m ^A = x^\m \sigma _\m^{A \dot A} \p _{\dot A}$, as follows
$ ( \l_1, \l_2, Y, 1) =
(\m^0,\m^1,\p_{\dot 0},\p_{\dot 1})/\p_{\dot 1}.$
The eq.(3.2) takes the form $K\p _{\dot A} =0$
and eq.(3.20) can be written in twistor terms as $K\m ^{0 A} = K\m ^A $.
This equation can be obtained also from $K \tau=1$ and  from the relations
$$K \m ^A = \dot x_0^\m \d_\m x^\n \sigma _\n^{A \dot A} \p_{\dot A} =
 \dot x _0^\n \sigma _\n^{A \dot A} \p_{\dot A}, $$
$$K \m ^{0 A} = \dot x_0^\m \d_\m x_0^\n \sigma _\n^{A \dot A} \p _{\dot A} =
 \dot x _0^\m \dot x_0^\n  \d _\m (\tau) \sigma _\n^{A \dot A} \p _{\dot A} =
 \dot x _0^\n  \sigma_\n^{A \dot A} \p_{\dot A} K\tau. $$}
$$ K \l_1^0 = K \l_1;\qquad  K \l_2^0 = K \l_2 .\eqno (3.20)$$ On the "left"
null plane we can use (3.10) and express $\D$ in the form $$ \D = (u-u_0)
\`v_0 + (\z - \z_0 ) \`\Z + (\Z - \Z_0) \`\z_0 + (v - v_0) \`u_0 = - \d_\t
(x - x_0 )^2 /2 = \t - t + \b V \b R , \eqno(3.21)$$ which gives a
retarded-advanced time equation $$\t = t \mp PZ^{-1} + \b V \b R,
\eqno(3.22)$$ and a simple expression for the solutions
$$ Y_1 =  [ (u-u_0)
\`v_0 + (\z - \z_0) \`\Z_0]/ [ (v - v_0) \`\Z_0 - (\Z -\Z_0) \`v_0],$$
 $$Y_2 =[ (v - v_0) \`u_0 + (\Z -\Z_0) \`\z_0 ]/ [(\Z -\Z_0) \`v_0 - (v -
v_0) \`\Z_0 ] . $$ Only the first root $Y_1$ is compatible with the
constraints (3.10) and satisfies the condition
$Y,_2=0$.
\bigskip
{\bf 4. NONSTATIONARY CASE. CONGRUENCES GENERATED BY A COMPLEX
WORLD LINE}\par
\bigskip
Now we would like to extend the above representation of the function F
in the form (3.3), and
introduce the "left" retarded time parameter $\t_L$  for
the nonstationary generalization of the Kerr congruence when $x_0(\t)$ is
an arbitrary complex world-line parametrized by complex time parameter $\t$
(not only straight and so far not only analytical).
For every point $x$ of the space-time one can consider the complex light
cone and the point $x_0(\t)$ of
the intersection of this light cone  with the complex world line.
One may look for a solution for the
parameter $\t$ of the corresponding light cone equation
$(x^\m - x^{0 \m})\eta_{\m\n}(x^\n-x_0^\n) = 0 $
where the metric $\eta_{\m\n}$ is (1.6), and the complex continuation
of the transformation (1.7) between null coordinates $x^{0\m}$ and
Cartesian coordinates holds. Using the complex Euclidean space distance
$r=\sqrt {(x-x_0)^2+(y-y_0)^2+(z-z_0)^2} $
between the real point $x$ and the complex point $x_0(\t)$,
and parametrisation $\t=t_0$,
we split this as
 $$\t = t - r (x, x_0 (\t))  \eqno(4.1). $$
This is an implicit nonlinear
equation for the retarded time coordinate $\t$. Its solution $\t(x)$ is
a complex scalar function on the space-time. One can introduce $K(\t) =
\dot x_0^\m(\t) \d_\m$ and, by the same method as in stationary case,
one  finds  that $$K(\t) \t(x)  = 1; \qquad K \l_1^0 = K
\l_1,\qquad K \l_2^0 = K \l_2 .\eqno (4.2)$$
We can find now the conditions on
the function $ F $ defined by (3.4), which will guarantee that its
solution Y satisfies the differential equation (2.7) for the shear free
and geodesic congruence. Differentiation (4.1) and
comparison the result with (2.7) yields the following equations $$
\t,_2=0  ; \eqno (4.3)$$ $$(\l_1 - \l_1^0) \d_\t K \l_2 - (\l_2 -\l_2^0)
\d_\t K \l_1 = 0 .\eqno (4.4)$$ To satisfy the equation (4.3) we attempt
to use a retarded time $\t$, which is subject to the light cone
constraint
$$(x_\m - x_{0 \m} (\t))(x^\m -x_0^\m(\t)) = 0.\eqno (4.5) $$
Differentiation of (4.5) gives $$ \t,_2 =[2 (x - x_0) e^1 + \bar\t,_2
\d_{\bar\t} (x - x_0)^2  ]/ \d_\t [(x - x_0)^2]  ,  \eqno (4.6)$$
and condition (4.3) takes the form
$$  (x - x_0) e^1 = 0; \eqno (4.7)$$ $$ \d_{\bar\t} x_0(\t) = 0.
\eqno (4.8)$$
Equation (4.4), in consequence of (4.7) and the representation
(3.9) for the twistor parameters, gives $$ (x -x_0) e^3 =0;
\eqno (4.9)$$ But (4.7) and (4.9) are in fact the equations (3.10) of
the "left" null plane. Thus the necessary conditions, under which the
congruence is shear free and geodesic,  are:
\par
  i) the world-line $ x_0(\t)$ must have an analytical dependence on
 the complex time $ \t,$ or
$ \d_{\bar\t} x_0(\t) = 0 ,$
\par
  ii) $\t$ must be a "left" solution of the retarded
 time equation - $\t_L$, corresponding to an intersection of the "left" null
plane with the world line $x_0(\t)$.
\par
 From (3.4) and (3.9) one can find
$$K \l_1 = {\dot x_0}^\m e^1_\m , \qquad K \l_2 = {\dot x_0}^\m
(e^3_\m - \Y e^1_\m), \eqno(4.10)$$
then by using (2.8), (2.9) and (4.1) we obtain
$$P = {\dot x_0}^\m e^3_\m , \qquad P_{\Y}=\partial _{\Y} P =
{\dot x_0}^\m e^1_\m , \qquad  Y,_3 = - Z P_{\Y}/P.
\eqno(4.11)$$
\bigskip
{\bf 5. ANALYSIS OF THE FIELD EQUATIONS }
\bigskip
We have not been able to obtain non-trivial real vacuum or
electrovacuum solutions for the above nonstationary generalization of the
Kerr metric.  However, we will present some results on the features of
such solutions and  show the existence of the complex radiating solutions,
having a stress-tensor proportional to $ e^3_\mu e^3_\nu $, which
included
the accelerating Kinnersley solution for the
photon rocket [3,8]
as particular case.  We shall assume that there is no electromagnetic field
and the null radiation may be produced by an incoherent flow of the light-like
particles like the case of the Kinnersley solution.
\par
First note that all equations for the Kerr-Schild form of metric of
the ref.[1] up to Eq.  (5.50) remain valid for a nonstationary case
if all variables are expressed in terms of the real function $P$.
However, for the nonstationary generalization of the Kerr
congruence considered above the function $P$ takes complex values,
that leads to the necessity of introducing corrections for three cases,
which we are going to consider.
\par
The function $P$ may be defined via  tangent directions of the complex
world line (4.11) and has to be essentially complex for any non-trivial
complex analytical world line.  Correspondingly, the function
$P (Y,\bar Y ,\tau)= \dot x_0^{\m} (\tau) e^3_{\m}$ has to be complex in
the general case too.
Integration of the field equations gives then
the complex meanings of $M$ which are not compatible with the standard
propositions of the real Kerr-Schild formalism [1].  However, we can obtain
that the Kerr-Schild formalism admits the complex solutions of the Einstein
equations.
\par
Independently of the real or complex meaning of the metric we have
$$R_{24} =R_{22} =R_{44}=0.  \eqno(5.1)$$
If the electromagnetic field is zero we have also
 $$R_{12} =R_{34}=0.   \eqno(5.2)$$
We mention that the equation
$$h,_{44} + 2(Z + \cZ)h,_4 + 2Z \cZ h =0,
\eqno(5.4)$$
which  follows from (5.2), admits not only real but also the complex
solutions $$ h= (MZ+ \bar M \cZ)/2  ,\eqno(5.5)$$
where $M$ and $\bar M$ are complex functions,
not necessarily complex conjugates,
obeying the conditions ${\bar M},_4 = M,_4=0 $.
\par
  Next, the equation $$ R_{23} = 0 , \eqno(5.6) $$ for the
complex $M$ acquires the form
$$ M,_2 - 3 Z^{-1} \cZ Y,_3 M - (Z^{-2}/2) [(M-{\bar M}) Z \cZ],_2 = 0
,\eqno(5.7)$$
containing the extra term
$$ (Z^{-2}/2) [(M-{\bar M}) Z \cZ],_2 .
\eqno(5.8)$$
  The last gravitational field equation
$R_{33} = -P_{33}$ takes the form
$$ M,_3 - Z^{-1} Y,_3 M,_1 - \cZ ^{-1} \Y ,_3 M,_2 =
Z^{-1} \cZ ^{-1} P_{33}/2  , \eqno(5.9) $$
where -$P_{33}$ is proportional to $e^3_{\m} e^3_{\n}$  that  corresponds to
the presence of null radiation.
\par
The exact solutions  we are looking for
satisfy the restriction $$M = \bar M  , \eqno(5.10)$$  leading to the
canceling of the extra term (5.8). As result the equation (5.7) acquires
the simple form and we get its complex solution $$ M= m/P^3, \eqno(5.11)$$
where $$m,_4 = m,_2 =0. \eqno(5.12)$$
 From (5.10) we have
$$ P= \bar P . \eqno(5.13)$$
This is possible for the complex functions $P$ and $M$ only if $\bar P$ and
$\bar M$ are considered as independent functions.
Thus, the complex world lines $x_0 ( \tau)$ and $ \bar x_0 ( \bar \tau)$ are
not to be taken as complex conjugated.  They may be considered rather to be
parallel, since their tangent vectors coincides, $\partial _{\tau}
x_0 = \partial _{\bar \tau} {\bar x_0} $ to provide the condition $$P =
\partial _{\tau} x_0^{\m} e^3_{\m}= \bar P = {\partial _{\bar \tau} {\bar
x_0}^{\m}}{\bar e^3_{\m}}.\eqno(5.14)$$
We have to speak now about the "right" and "left" world lines
$x_0$ and $\bar x_0, $ , about the corresponding "right" and "left"
time parameters $\tau_L$, $\bar \tau_R$ and parameters  $Y$ , $\Y$.
\par
We introduce the differential operator
$$ {D} = \partial _3 - Z^{-1} Y,_3 \partial _1  - \cZ ^{-1} \Y ,_3
\partial _2 . \eqno(5.15) $$
Action of the operator $D$ on the variables $Y, \bar Y $ and $ \tau$
is following
$$ D Y = D \bar Y = 0,\qquad  D \tau = P^{-1}, \eqno(5.16)$$
where the last relation follows from (4.2) and (4.11).
If $ M$ is function of $Y, \bar Y$ and $ \tau$ then the equation (5.9)
takes the form
    $$ \partial_\tau M = P  Z^{-1} \cZ ^{-1} P_{33}/2  , \eqno(5.17) $$
It is not really a field equation but a definition of $P_{33}$ of
the null radiation. Substitution of (5.11) gives
   $$ P_{33} = 2 P Z \cZ  \partial_\tau m/P^3
=  Z \cZ  [-6m(\partial_\tau P) + 2 P(\partial_\tau m)]/P^3, \eqno(5.18) $$
The resulting metric is complex and has the form
 $$g_{\m\n} =
\h_{\m\n} + (m/P^3)(Z +\bar Z) e^3_{\m} e^3_{\n}. \eqno(5.19) $$
One can normalize $e^3$ by introducing $l = e^3/P$ so that
$${\dot x}_0^\m e^3_\m =1, \eqno(5.20) $$ and the metric takes the
form
 $$g_{\m\n} =\h_{\m\n} - m({\tilde r}^{-1} + \bar{\tilde r}^{-1})
l_{\m} l_{\n}, \eqno(5.21) $$
where $\tilde r$ is a complex radial coordinate
$$ \tilde r =  PZ^{-1} = - dF/dY . \eqno(5.22) $$
We can select also two particular cases.
The first case corresponds to the known nonstationary
solutions of Kinnersley [8] where congruence is twist free.
The world line is real in this case,
$ Im x_0 = 0 $  and $ \tau $ is real too. The  complex
conjugate world lines coincide and so do the "right" and "left"
retarded times.
\par
Equation (5.11) has the solution $$ M (Y, \bar Y, u) = m/P^3,
\eqno(5.19) $$ where function $P$ depends now on real $u=\tau= \bar \tau$
and $m,_4 = m,_2 =0. $ From the equation (5.18) one can see that
metric is not vacuum.  We get the real retarded-time construction considered
by Kinnersley [8] leading to a generalization of the Vaidya shining-star
metric that permits arbitrary acceleration of the source.
The Kinnesley metric has the Kerr-Schild form
 $$g_{\m\n} = \h_{\m\n} + H l_\m l_\n =
\h_{\m\n} + 2(m/r) (\sigma_\m /r) (\sigma_\n /r). \eqno(5.23) $$
The relation with our notations is following
$$l^\m = \sigma^\m/r, \eqno(5.24) $$
where $$\sigma^\m = x^\m - x_0^\m,\qquad r = PZ^{-1}. \eqno(5.25)$$
The Kinnersley retarded time parameter $ u = \tau/ \sqrt {(\dot x_0)^2},$
and $\lambda^\m (u) = \dot x_0^\m (\tau)/ \sqrt{(\dot x_0)^2}.$
The metric  has a Ricci tensor
proportional to $e^3_{\m} e^3_{\n}$  that  corresponds to the presence of
null radiation.
\par
The second degenerate case corresponds to a straight complex world line
having a constant complex tangent direction $\xi = \partial _{\tau} x_0 =
const.$.
The function $P= \xi^{\m} e^3_{\m} $ is now independent of $\tau$.
The equation (5.17) is fulfilled with $m=const, P_{33}=0$  and we get
the exact  complex vacuum solution which is generalization of the Kerr
solution to the case of a complex Killing direction $\xi^\m$.
The physical meaning of this solution is still unknown and may be obtained
only after finding the real slice.
\bigskip
{\bf CONCLUSION}\par
\bigskip
We describe  nonstationary geodesic shear free principal null
congruences in  Kerr-Schild spaces which are generated by a complex world
line.
Similar ideas were considered also in papers [4-7] and in the case
of a real world line  in work [8] where an exact solution was pointed out.
  The present consideration in the Kerr-Schild formalism gives a
convenient form of the function F, of the Kerr theorem used in
the retarded time construction and allows us to obtain an explicit
representation of the congruence and singularities. We make precise also
the necessary conditions on the complex world line and retarded time
parameter.
The analysis of field equations shows that there are complex and radiating
metrics among the nonstationary generalizations of the Kerr solution.
\par
The second feature is an appearance  of the broken
symmetry of complex conjugation  and a possibility of the equivalent
description of this situation in the class of double Kerr-Schild
metrics [9,10].
\par
 Complex world lines may be considered in string theory as a particular
class of relativistic strings [11,12]. A further application of the complex
world line representation of shear free geodesic congruences has been pointed
out in [12].
\par
In conclusion we would like to thank C.B.G. McIntosh for stimulating
discussions.
\bigskip
\vfill
\eject
\newpage
{\bf REFERENCES}
\frenchspacing
\medskip
\baselineskip=15.5pt
\item{[1]}  G.C. Debney,R.P. Kerr,
A.Schild,J.Math.Phys.,{\bf10}(1969) 1842.
\item{[2]}  R. Penrose,J. Math. Phys.,{\bf8}(1967) 345.
\item{}
  R. Penrose, W. Rindler, Spinors and space-time.v.2.
Cambridge Univ. Press, England,1986.
\item{}
D.Cox and E.J. Flaherty, Commun. Math. Phys. {\bf 47}(1976)75.
\item{[3]} D.Kramer, H.Stephani, E. Herlt, M.MacCallum, Exact Solutions
 of Einstein's Field Equations, Cambridge Univ. Press, Cambridge 1980.
\item{[4]} R.P. Kerr, W.B. Wilson, Gen. Rel. Grav.,{\bf10}(1979)273.
\item{[5]}  A.Ya. Burinskii, in: Problem of theory of gravitation and
   elementary particles,{\bf11}(1980), Moscow,Atomizdat,(in russian).
\item{}
   D. Ivanenko and A.Ya. Burinskii, Izvestiya Vuzov Fiz. $N^0$ 7 (1978)
   113 (Sov. Phys. J. (USA)).
\item{[6]}  E.T. Newman,J.Math.Phys.,{\bf14}(1973)102.
\item{[7]} R.W. Lind, E.T. Newman,J. Math. Phys.,{\bf 15}(1974)1103.
\item{[8]} W. Kinnersley, Phys. Rev. {\bf 186} (1969) 1335.
\item{[9]} J.F.  Plebanski  and  A.  Schild, Complex Relativity and
Double KS Metrics,  in:   Proceedings   of   the
International Symposium on  Mathematical  Physics,  Mexico  City,
Mexico, 5-8 January 1976,pp 765-787.
\item{[10]}J.F. Plebanski and I. Robinson, Phys. Rev. Letters,
{\bf 37}(1976)493
\item{[11]} W.T. Shaw, Class. Quant. Grav.{\bf 3}(1986)753
\item{[12]} A.Ya. Burinskii, String-like Structures in Complex Kerr
Geometry, in: Relativity Today, Proccedings of the Fourth Hungarian
Relativity Workshop, Edited by R.P.Kerr and Z.Perj\'es,
Akad\'emiai Kiad\'o, Budapest,1994 p.149; gr-qc/9303003
\item{}
 A.Ya. Burinskii, Phys.Lett. {\bf A 185} (1994) 441
\end
\end